# Control of the magnetism and magnetic anisotropy of a single-molecule magnet with an electric field

*Jun Hu and Ruqian Wu*

*Department of Physics and Astronomy, University of California, Irvine, California 92697-4575, USA*

**Abstract.** Through systematic density functional calculations, the mechanism of the substrate induced spin reorientation transition in FePc/O-Cu(110) was explained in terms of charge transfer and rearrangement of Fe-3d orbitals. Moreover, we found giant magnetoelectric effects in this system, manifested by the sensitive dependence of its magnetic moment and magnetic anisotropy energy on external electric field. In particular, the direction of magnetization of FePc/O-Cu(110) is switchable between in-plane and perpendicular axes, simply by applying an external electric field of 0.5 eV/Å along the surface normal.

Manipulation of magnetic properties of nanomaterials with an external electric field (EEF) through the magnetoelectric effect is extremely attractive for the development of both fundamental science and innovative spintronics devices. [1,2] The magnetoelectric responses of nanomaterials are typically much enhanced with respective to their bulk counterparts due to the size reduction, quantum confinement effect and weakened screening. For instance, the magnetic ordering of a Mn-Mn dimer on the Ag(001) surface can be conveniently switched between the ferromagnetic and antiferromagnetic states, by using an electric field of ~0.5 V/Å. [3] In particular, extensive studies have been devoted to establish fundamental understanding for how to control the magnetic anisotropy with EEF, since the orientation of magnetization is of high importance for applications of nanomagnets. [4,5,6,7,8,9] As building blocks in innovative spintronics and molecular electronics nanodevices, organic magnetic molecules are of special research interest. [10,11] It was found that the easy axis of magnetization of Fe-phthalocyanine (FePc) molecules turns from the in-plane direction to the perpendicular direction in touch with the oxidized Cu(110) [O-Cu(110)] surface. [12] For the development of molecular spintronics, it is critical to establish clear insights for the substrate-induced spin reorientation transition (SRT) and, furthermore, the magnetoelectric effect on a prototype magnetic molecular system such as FePc/O-Cu(110).

In this Letter, we report results of density functional theory (DFT) calculations for the electronic and magnetic properties of FePc/O-Cu(110). The mechanism of the substrate-induced SRT was revealed, using the energy-level shifts and spin-orbit coupling (SOC) matrices of molecular orbitals. It is striking that both magnitude and sign of the magnetocrystalline anisotropy energy ($E_{MCA}$) of FePc/O-Cu(110) can be altered by a moderate EEF, because of the electric field-induced electron charge transfer between the FePc molecule and the substrate. Our findings indicate that FePc/O-Cu(110) is a promising model magnetoelectric system for fundamental studies and spintronics applications.

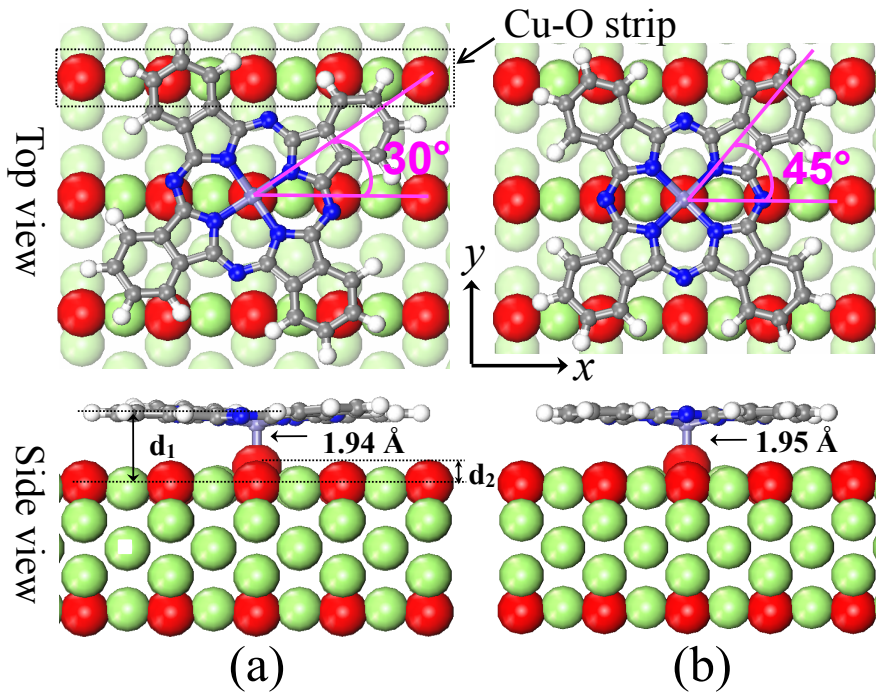

*Figure 1. (Color online) (a) The α/O and (b) the β/O adsorption geometries of FePc/O-Cu(110) with both top- and side-views. The lower panels also show the optimized Fe-O bond lengths. Red and green spheres are for the oxygen and Cu atoms of the substrate. To make the surface Cu atoms more distinguishable, Cu atoms in subsurface layer are represented by light green spheres in the upper panels.*

DFT calculations were carried out with the Vienna ab-initio simulation package (VASP), [13,14] at the level of the spin-polarized generalized-gradient approximation (GGA). [15] To examine the reliability of structural models and electronic properties, the non-local van der Waals density functional (vdW-DF) which may significantly improve the adsorption of large molecules, [16,17] and Hubbard U correction (GGA+U) which accounts for the strong on-site Coulomb interactions among 3d electrons, [18] were also used. We used the projector augmented wave (PAW) method for the description of the ionic cores. [19,20] As sketched in Fig. 1, the O-Cu(110) substrate was simulated by a slab model that has three Cu layers and one Cu-O overlayer on each side, along with a 15 Å vacuum between adjacent slabs. The FePc molecule was placed on O-Cu(110) in different angles, with the Fe-N axis along either 30° (denoted as the α-type geometry) or 45° (denoted as the β-type geometry) away from the [001] axis of the Cu lattice. [12] In addition, the core Fe atom of the molecule may take site above either Cu or O atom of the substrate. Therefore, we considered four possible geometries, referred as α/Cu (α-type, Fe on Cu), β/Cu (β-type, Fe on Cu), α/O (α-type, Fe on O), and β/O (β-type, Fe on O). Nevertheless, the adsorption sites were not constrained since the molecule was allowed to shift sideways in calculations. To mimic adsorption of the single FePc molecule, a large 6×3 supercell in the lateral plane was adopted, with a dimension of 18.07×15.34 Å$^2$. The energy cutoff for the plane wave expansion was 400 eV, adequate for FePc/O-Cu(110) according to our test calculations. A 3×3 k-grid mesh was used to sample the tiny tow dimensional Brillouin zone. The bottom CuO layer and one Cu layer were fixed, while the atomic positions in other layers were fully relaxed using the conjugated gradient method for the energy minimization procedure, with a criterion that requires force on each atom smaller than 0.01 eV/Å.

Scanning tunneling microscope (STM) experiments established that oxygen atoms take the long bridge sites over the Cu(110) surface to form the striped CuO overlayer. [21,22,23] Our total energy calculations confirmed that this is indeed the ground state geometry of the O-Cu(110) substrate. Furthermore, we found that the Cu-O rows ripple on O-Cu(110), with O atoms higher than Cu atoms by 0.15 Å [see Figs. 1(c) and (d)], in good accordance with the experimental data, 0.21±0.1 Å. [21] The non-vanishing density of states at Fermi level of the CuO stripe indicates that the CuO stripe is actually metallic, although it is assumed to act as insulating layer to separate the adsorbates and the Cu(110) substrate. [12,24]

The stability of FePc/O-Cu(110) in different geometries is characterized by the binding energy that is defined as

$$E_b = E[\text{O-Cu(110)}] + E(\text{FePc}) - E[\text{FePc/O-Cu(110)}] . \quad (1)$$

Here, E(FePc) is the total energy of the free FePc molecule, while $E[\text{FePc/O-Cu(110)}]$ and $E[\text{O-Cu(110)}]$ stand for the total energies of O-Cu(110) with and without the presence of the FePc molecule, respectively. As listed in Table I, the FePc molecule prefers the α/O geometry on O-Cu(110) since it has the largest $E_b$ (0.36 eV) among all four configurations. Although the magnitude of $E_b$ is small, the FePc molecule deforms remarkably and, on the other hand, causes a significant surface reconstruction on the O-Cu(110) substrate. As displayed in the bottom panels of Fig. 1, the O atom right under Fe is pulled out of the CuO stripe by as much as 0.7 Å (denoted as $d_2$). On the other side, Fe and its four N neighbors in the FePc molecule drop down from the molecular base plane by 0.6 Å and 0.3 Å, respectively. Similar structural deformation was also reported for Sn-Phthalocyanine molecule adsorbed on Ag(111) surface, where the central Sn atom is pulled down by ~0.5 Å towards the substrate. [25] As a result, the Fe-O bond length is only 1.94 Å, indicating a strong attraction between the two atoms. On the contrary, the carbon rings and substrate repel each other, so $d_1$ is as large as 3.1 Å. Therefore, FePc/O-Cu(110) manifests mixed features of chemisorption and physisorption, i.e., a strong ionic bond but with a very small binding energy.

***Table I.*** *Binding energy ($E_b$), total spin moment ($M_S$), magnetocrystalline anisotropy energy ($E_{MCA}$) and geometry parameters of the FePc molecule in the free space and on the O-Cu(110) surface. Values in parentheses are calculated with the vdW-DF*

*correction. Note that values of $d_1$ and $d_2$ are averaged over the deformed FePc molecule on O-Cu(110).*

|  | Free | α/Cu | β/Cu | α/O | β/O |
|---|---|---|---|---|---|
| $E_b$ (eV) |  | 0.19 (0.67) | 0.10 (0.60) | 0.36 (0.86) | 0.33 (0.83) |
| $d_{Fe-O}$ |  |  |  | 1.94 (1.95) | 1.95 (1.95) |
| $d_{Fe-Cu}$ |  | 3.14 (3.57) | 3.14 (3.54) |  |  |
| $d_1$ |  | 3.1 (3.6) | 3.2 (3.7) | 3.1 (3.5) | 3.2 (3.4) |
| $d_2$ |  | 0.1 (0.2) | 0.3 (0.2) | 0.8 (1.0) | 0.8 (1.0) |
| $M_S$ ($\mu_B$) | 2.00 | 1.84 (1.96) | 1.69 (2.00) | 2.40 (2.54) | 2.43 (2.52) |
| $E_{MCA}$ (meV) | -1.24 | -0.93 (-1.17) | -0.93 (-1.15) | 0.48 (0.23) | 0.46 (0.53) |

Note that α/Cu and β/Cu geometries were assigned as the ground state geometries in early report [12], different from what we found here through GGA calculations. To solve this puzzle, we also optimized all four structures with the vdW-DF and GGA+U approaches. The inclusion of Hubbard correction with U up to 4 eV appears not to change atomic structure but the vdW-DF correction noticeably affects the atomic structure. As given in parentheses in Table 1, vdW-DF calculations give larger $d_1$ and $E_b$ for all four cases, compared to GGA data. Nonetheless, neither vdW-DF nor GGA+U correction affects the adsorption site preference. For example, the energy difference between α/O and α/Cu geometries with the vdW-DF correction is 0.19 eV, very close to the corresponding GGA result, 0.17 eV. In addition, the energy differences between α and β geometries are not much affected either (0.03 eV on O and 0.07 eV on Cu). Therefore, we believe that the assignment of α/Cu as the ground state geometry was a mistake. In the following, we mostly focus on GGA results of the α/O geometry, with a note that properties of the co-existing β/O geometry are not much different.

According to Bader's charge analysis scheme, [26] the iron atom in FePc transfers 0.43 electrons to the oxygen atom underneath. As a result, the spin magnetic moment ($M_S$) of FePc/O-Cu(110) enhances to 2.40 $\mu_B$ compared to 2.00 $\mu_B$ in the freestanding case. This value agrees excellently with the experimental data, 2.30±0.02 $\mu_B$. [12] Significant spin-polarization is induced around the O and Cu atoms adjacent to Fe, with $M_S$ of 0.15 $\mu_B$ for O and 0.03 $\mu_B$ for Cu, respectively. To better appreciate the molecule-substrate interaction, we plot the partial density of states (PDOS) in Fig. 2 for both the free and the supported FePc molecules. Since the freestanding FePc molecule has a $D_{4h}$ symmetry, the Fe-3d orbitals split into four groups: $b_{1g}$ (xy) and $b_{2g}$ ($x^2$-$y^2$) for the in-plane components, along with $a_{1g}$ ($z^2$) and $e_g$ (xz and yz) for the out-of-plane components. [27,28] However, the actual energy spectrum of the Fe-3d orbitals in Fig. 2(a) is somewhat different from this simple assignment, because of the interaction with N-2p states. [29] For example, the lowest peak in the minority spin channel comprises of $b_{2g}$, $e_g$ and $a_{1g}$ features all together; the peak right above the Fermi level combines the $b_{2g}$ and $e_g$ components. When the FePc molecule is placed on O-Cu(110), the Fe-$a_{1g}$ orbital becomes delocalized as shown by the broad PDOS features in Fig. 2(b). This manifests the strong hybridization between the Fe-$a_{1g}$ and O-$p_z$ orbitals. Significantly, the two PDOS peaks across the Fermi level become more "pure": with the $b_{2g}$ state below $E_F$ and the $e_g$ state above $E_F$. As a result, the contour plot of the charge density difference shows the intra-molecular charge transfer from the Fe-$e_g$ orbital to the Fe-$b_{2g}$ orbital in Fig. 2(c).

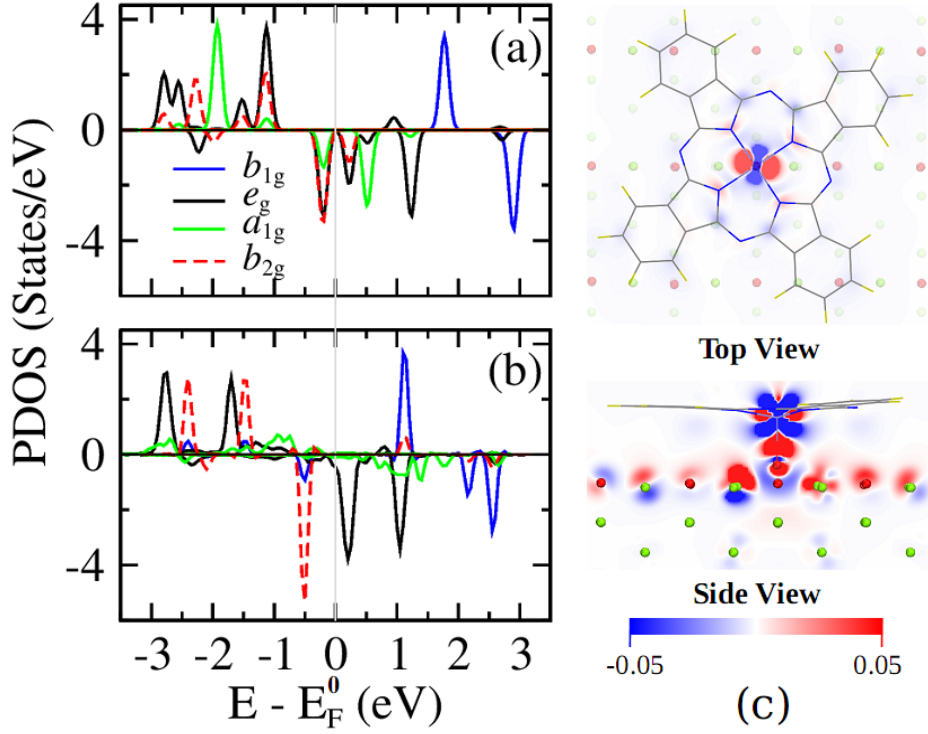

***Figure 2.*** *Partial density of states (PDOS) of Fe-d orbitals in (a) the freestanding FePc molecule, and (b) FePc/O-Cu(110). Positive and negative PDOS are for the majority and minority spin channels, respectively. The gray vertical line at E=0 indicates the position of $E_F$. (c) Electron density difference: $\Delta\rho=\rho[FePc/O-Cu(110)] - \rho[O-Cu(110)] - \rho(FePc)$. Blue and red regions show charge depletion and accumulation, respectively.*

To determine the magnetocrystalline anisotropy energy, we adopted the torque approach proposed by Wang et al., [30,31]

$$E_{MCA} = \sum_{i \in occ} \left\langle \psi_i \left| \frac{\partial H_{SO}}{\partial \theta} \right| \psi_i \right\rangle_{\theta=45°} \quad (2)$$

Here, $\Psi_i$ is the $i^{th}$ relativistic eigenvector, and $H_{SO}$ is the SOC Hamiltonian. We recently implemented this approach in the framework of VASP, by transforming the SOC operator to [19]

$$\widetilde{H}_{SO} = \sum_{m,n} |\tilde{p}_m\rangle \langle \phi_m | H_{SO} | \phi_n \rangle \langle \tilde{p}_n |. \quad (3)$$

Here, $\tilde{p}_m$ and $\phi_m$ are the projector functions and all-electron partial waves in the augmentation region as used in the PAW method. As a benchmark test for the new implementation, the calculated $E_{MCA}$ of a free FePc molecule is -1.24 meV, in good agreement with that obtained from the all-electron full-potential linearized augmented plane wave method, -1.18 meV. [28] The negative sign of $E_{MCA}$ indicates that the easy axis lies in the base plane of the molecule, in good accordance with experimental observations. [32,33] As seen in Table I, $E_{MCA}$ of FePc/O-Cu(110) indeed changes to positive, 0.48 meV, so the switch of the easy axis to the perpendicular direction reported by Tsuhakara *et al* [12] is confirmed. Although the amplitude of $E_{MCA}$ is somewhat changed by vdW-DF or GGA+U correction (c.f., data in Table 1 and in Supplemental materials), the substrate induced spin reorientation transition is unaffected. Interestingly, $E_{MCA}$ remains negative for FePc on top of Cu, in GGA, GGA+U and vdW-DF calculations. This is another evidence that FePc takes the O site rather than the Cu site on O-Cu(110) as claimed before. [12, 34] It is worthwhile to point out that Ref. [34] claimed good agreement with the experimental inelastic electron tunneling spectroscopy (IETS) without attempting magnetic anisotropy calculations. The "agreement" results from experimental parameters [12], and is hence useless for the assignment of preferential adsorption site.

Now we can explore for the reasons that cause the substrate-induced SRT in FePc/O-Cu(110) and thenceforth can find out ways to control it. Following the second order perturbation approach proposed by Wang, Wu and Freeman, [35] $E_{MCA}$ can be approximately determined by matrix elements of the angular momentum operators: $L_z$ and $L_x$, across the unoccupied (u) and occupied (o) states,

$$E_{MCA} \approx \xi^2 [\sum_{u,o} \frac{<u|L_z|o>^2}{E_u - E_o} - \frac{<u|L_x|o>^2}{E_u - E_o}]. \qquad (4)$$

Similar procedure was also discussed for the calculations of magnetic anisotropy of molecular magnets. [36] For convenience of analysis, we further subdivide contributions from the majority spin states [$E_{MCA}$(uu)], the minority spin states [$E_{MCA}$(dd)], and also the cross-spin coupling [$E_{MCA}$(ud+du)]. For simplicity, we discuss $E_{MCA}$(dd) in details because it contributes most part of the total $E_{MCA}$ for both free FePc and FePc/O-Cu(110) as discussed later. In Figs. 3(a) and 3(b), we constructed a simple energy spectrum of Fe-*3d* orbitals in the minority spin channel and plot all non-vanishing $L_z$ and $L_x$ elements across these states. With this construction, one may easily estimate $E_{MCA}(dd)$ by inspecting the number and weight of lines that intercept the Fermi level.

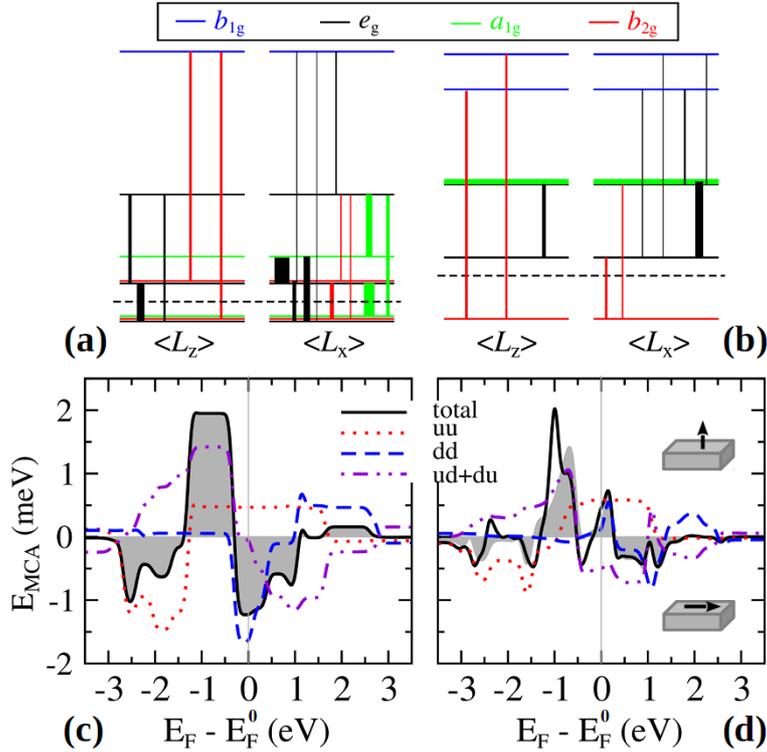

**Figure 3.** *(a) and (b) Sketches of energy spectrum of Fe-3d states in the minority spin channel and non-vanishing SOC matrix elements for the free FePc molecule and the FePc/O-Cu(110) system, respectively. The horizontal dashed line shows the position of the actual Fermi level for each case, $E_F^0$. The thickness of each vertical line scales with the magnitude of the corresponding SOC matrix element, and its color matches to the wave function feature of the lower state in the pair. (c) and (d) Total and spin-decomposed $E_{MCA}$ of the free FePc molecule and the FePc/O-Cu(110) system. Shaded regions show the total $E_{MCA}$ with SOC contributions solely from Fe.*

As clearly shown in Fig. 3(a), in the critical SOC pairs of the occupied states to the empty states there is an imbalance between the seven $L_x$ contributions to the three $L_z$ contributions, which leads to a negative $E_{MCA}(dd)$ for the free FePc molecule. In FePc/O-Cu(110), the $a_{1g}$ orbital become delocalized and the $e_g$ orbital shifts to the unoccupied region [see Figs. 2(b) and 3(b)]. As a result, two thin $L_x$ lines (i.e., with small $L_x$ elements) and two thick $L_z$ lines (i.e., with large $L_z$ elements) intercept the Fermi level and $E_{MCA}$(dd) thereby becomes positive. Following this argument, the number of $L_x$ lines exceeds that of $L_z$ lines if its Fermi level shifts up to above the $e_g$ state as seen in Fig. 3(b), by adding excessive electrons into the molecule, $E_{MCA}$(dd) of FePc/O-Cu(110) may become negative again. Such analyses can establish trends of $E_{MCA}$(dd) with respect to a shift of the Fermi level, $E_F - E_F^0$, in order to guide an experimental search. Quantitatively, we directly calculated the total and spin-components of $E_{MCA}$ in a broad range of $E_F - E_F^0$ using the rigid band model. As shown in Fig. 3 (c), $E_{MCA}$(dd) of the free FePc molecule remains negative in the range -0.5 eV < $E_F - E_F^0$ < 1.1 eV. In contrast, $E_{MCA}$(dd) of FePc/O-Cu(110) changes sign quickly as seen in Fig. 3(d); it becomes negative at $E_F - E_F^0 \approx 0.4$ eV, where the Fermi level moves to above the $e_g$ state in the minority spin channel [cf. Fig. 2(b)]. Interestingly, $E_{MCA}$(uu) and $E_{MCA}$(ud+du) of FePc/O-Cu(110) cancel each other within the range -0.5 eV < $E_F - E_F^0$ < 1.0 eV, although their absolute amplitudes are even larger than that of $E_{MCA}$(dd). As a consequence, $E_{MCA}$(dd) plays the dominant role in the substrate-induced SRT. In addition, $E_{MCA}$ solely originates from the SOC effect of the Fe atom for the free FePc molecule, as suggested by the perfect overlap between bold-solid line (SOC contributions from all atoms were included) and the shaded region (SOC of only the Fe atom was included) in Fig. 3(c). The overlap becomes less perfect for FePc/O-Cu(110) in Fig. 3(d), because of the minor contributions from the SOC effect of Cu atoms in the substrate.

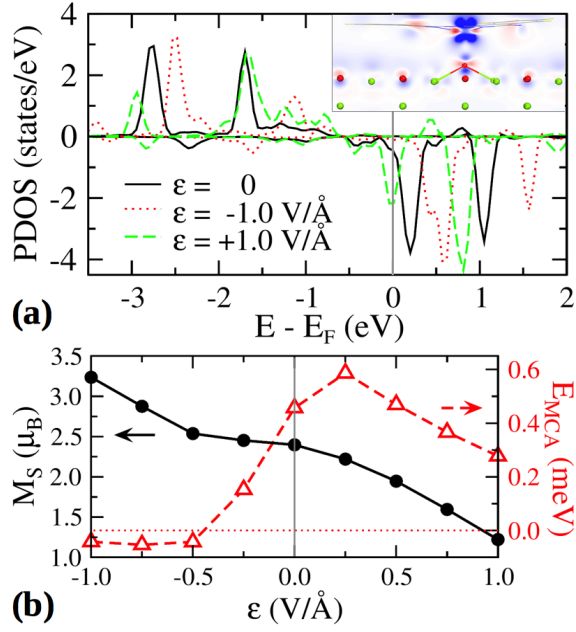

***Figure 4.*** *(a) PDOS of the Fe $e_g$ orbital of FePc/Cu(110) under different electric fields. A positive $\varepsilon$ is defined as pointing from the molecule to the substrate. The inset shows the field induced charge redistribution: $\Delta\rho = \rho(\varepsilon=-1.0\ V/Å) - \rho(\varepsilon=0)$. The atomic symbols and color scale of the charge density are the same as that in Fig. 2(c). (b) $M_S$ and $E_{MCA}$ of FePc/O-Cu(110) as a function of $\varepsilon$.*

Note that $E_{MCA}$ of FePc/O-Cu(110) changes rapidly near $E_F - E_F^0 = 0$, with a positive slope as shown in Fig. 3(d). This offers an opportunity to tune the magnetic anisotropy of FePc/O-Cu(110) by applying an electric field ($\varepsilon$). Here, we define the electric field pointing downward to the surface as positive. Since the screening in the region between the molecule and the substrate is rather weak, the electronic potential around the FePc molecule may easily shift to lower (higher) value with respect to the substrate by a positive (negative) EEF, and so do the PDOS peaks of Fe-*3d* orbitals. The electric field dependence of the Fe-$e_g$ peak was used as example in Fig. 4(a). It appears that the magnitude of the energy shift with $\varepsilon$ = -1.0 V/Å is larger than that with $\varepsilon$ = 1.0 V/Å. The presence of EEF alters the electron population of the FePc molecule and also its magnetic moment, as shown in Fig. 4(b). At $\varepsilon$ = +1.0 V/Å, the Bader charge of FePc molecule becomes -0.66e, compared to +0.43e for the zero-field case. When $\varepsilon$ is -1.0 V/Å, the Bader charge of FePc molecule is +1.39e. [37] The inset in Fig. 4(a) shows that a negative EEF causes charge depletion from the Fe-$e_g$ state to the substrate. The large range of $M_S$ in Fig. 4(b), from 1.15 $\mu_B$ at $\varepsilon$ = +1.0 V/Å to 3.24 $\mu_B$ at $\varepsilon$ = -1.0 V/Å, suggests a giant magnetoelectric effect in FePc/O-Cu(110).

It is interesting that the calculated $E_{MCA}(\varepsilon)$ curve in Fig. 4(b) closely follows the trend of $E_{MCA}(E_F - E_F^0)$ in Fig. 3(d), as predicted by the rigid band model analysis. On the positive side of $\varepsilon$, $E_{MCA}$ first increases to its summit at $\varepsilon$ =0.25 V/Å and then drops gradually afterward. For negative $\varepsilon$, $E_{MCA}$ decreases rapidly and changes its sign near $\varepsilon$ = -0.5 V/Å. This is caused by electron depletion from the Fe-$e_g$ orbital as well as by the involvement of Cu-d states as shown in the inset of Fig. 4(a). This EEF-induced SRT is very important for magnetic recording and spintronics applications since one has a means to switch the easy axis of FePc/O-Cu(110) between the in-plane and perpendicular direction.

In summary, structural, electronic and magnetic properties of the FePc molecule on the O-Cu(110) surface have been systematically studied through density functional theory calculations. We have shown that the FePc molecule forms a strong ionic Fe-O bond with the substrate, even though the adsorption energy is small. In this system, the charge transfer is the main cause for the substrate induced SRT, according to both second-order perturbation analysis and rigid band model calculations. Intriguingly, we found that the spin orientation of FePc/O-Cu(110) is switchable by applying a negative external electric field. Our studies pave a way for the mechanism-based design of molecular spintronics devices.

**Acknowledgements**

We thank Prof. W. Ho for insightful discussions. Work was supported by DOE-BES (Grant No: DE-FG02-05ER46237) and by NERSC for computing time.

# Control of the magnetism and magnetic anisotropy of a single-molecule magnet with an electric field

*Jun Hu and Ruqian Wu*

*Department of Physics and Astronomy, University of California, Irvine, California 92697-4575, USA*

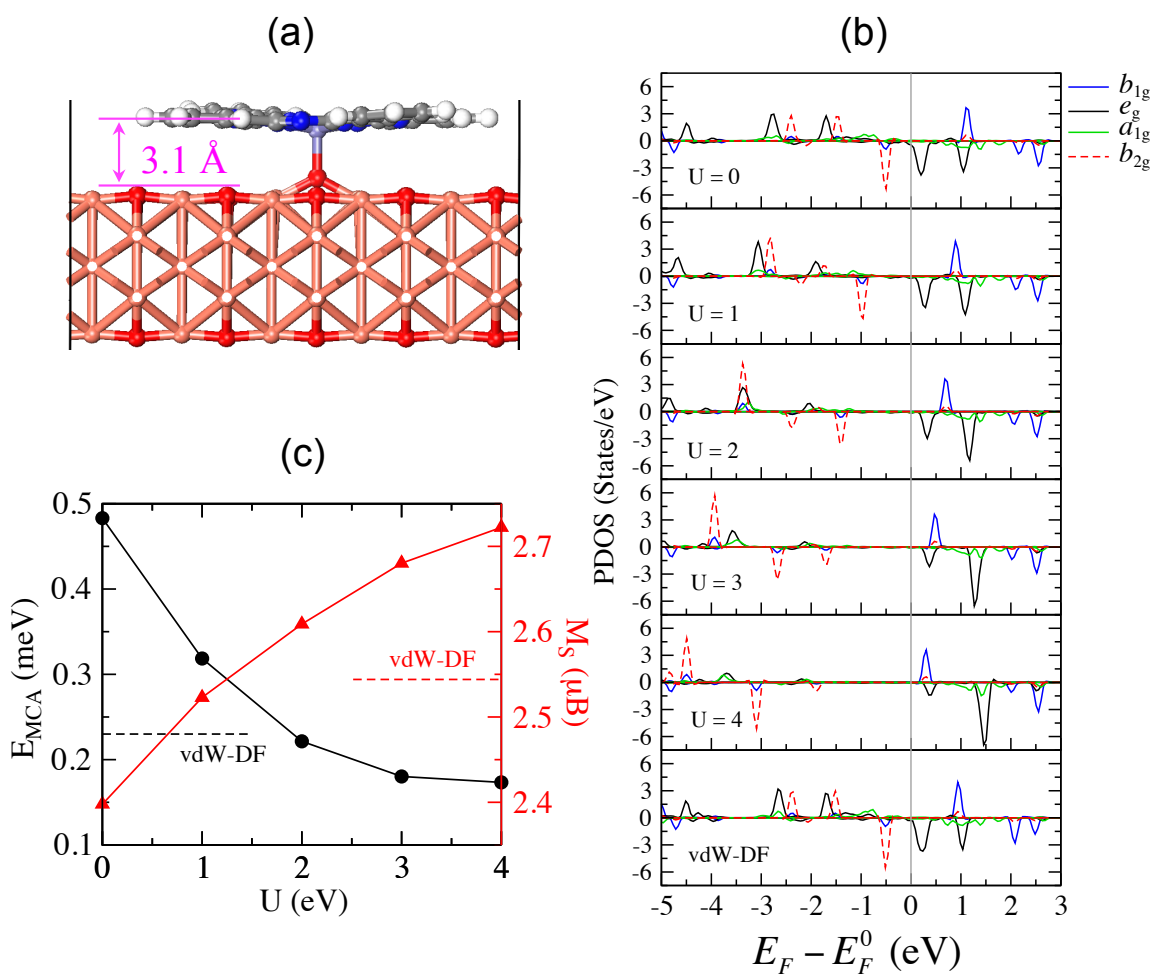

**Figure S1.** *(a) Side view of the α/O adsorption geometry of FePc/O-Cu(110) from GGA calculation. The coral, red, purple, blue, gray and white spheres stand for the Cu, O, Fe, N, C and H atoms. The average distance between the FePc molecular plane and the O-*

Cu(110) substrate is marked. (b) Projected density of states (PDOS) of Fe-3d orbitals with different U corrections as well as vdW-DF corrections. (c) The magnetocrystalline anisotropy energy ($E_{MCA}$) (black dots) and spin moment ($M_S$) (red triangles) of α/O FePc/O-Cu(110) as functions of U. The horizontal dashed lines indicate the corresponding values of vdW-DF calculations.

We explored the effect of Hubbard U correction on our results, with U=1-4 eV for the Fe-3d orbitals. The atomic structures were re-optimized but no notable change was found from GGA results. Furthermore, GGA+U calculations do not change the site preference of FePc on O-Cu(110). For example, the energy difference between α/O and α/Cu geometries with the GGA+U correction is 0.14 eV, very close to the corresponding GGA result, 0.17 eV.

The inclusion of U causes shifts of energy levels, as shown for the α/O geometry in Fig. S1(b). The occupied $b_{2g}$ and $e_g$ orbitals and unoccupied $b_{1g}$ orbital are pushed downwards, whereas the unoccupied $e_g$ orbital are pushed upwards. Furthermore, the electron occupancy of the $e_g$ orbital decreases as the value of U increases. As a result, the spin moment of FePc/O-Cu(110) increases monotonically with U, as seen in Fig. S1(c). Meanwhile, $E_{MCA}$ decreases in magnitude, due to the increases of the denominators in Eq. 4 in the text. Significantly, the sign of $E_{MCA}$ for the α/O geometry keeps positive for U up to 4 eV, so the substrate induced spin reorientation as discussed in the text is unaffected by the U-term.

Considering that the measured $M_S$ of FePc/O-Cu(110) is 2.30±0.02 $\mu_B$, [**Error! Bookmark not defined.**] we believe that GGA is more appropriate for the description of electronic and magnetic properties of FePc/O-Cu(110). The molecular orbitals are rather delocalized due to the intermixing with N-p orbitals. Therefore, the on-site Coulomb interactions should not significant in FePc/O-Cu(110) and the Hubbard U correction is actually unnecessary.